\documentclass[reprint,aps,prl,twocolumn,superscriptaddress]{revtex4-1}
\usepackage{braket}
\usepackage{amssymb}
\usepackage{amsmath}
\usepackage{epsfig}
\usepackage{color}
\usepackage{graphics, graphicx}
\usepackage{bbold}
\usepackage{psfrag}
\usepackage{mathcomp}
\usepackage{subfigure}
\usepackage{verbatim}
\usepackage[colorlinks, citecolor=blue]{hyperref}
\usepackage[normalem]{ulem}
\usepackage{bm}

\begin{document}

\title{Moiré droplet of ultracold Bose gases in a twisted-bilayer optical lattice}
\author{Tianqi Luo}
\affiliation{Beijing National Laboratory for Condensed Matter Physics, Institute of Physics, Chinese Academy of Sciences, Beijing 100190, China}
\affiliation{Department of Basic Courses, Naval University of Engineering, Wuhan 430033, China}
\author{Jing Zhang}
\affiliation{State Key Laboratory of Quantum Optics and Quantum Optics Devices, Institute of Opto-Electronics,
Collaborative Innovation Center of Extreme Optics, Shanxi University, Taiyuan, China}
\author{Xiaoling Cui}
\email{xlcui@iphy.ac.cn}
\affiliation{Beijing National Laboratory for Condensed Matter Physics, Institute of Physics, Chinese Academy of Sciences, Beijing 100190, China}

\date{\today}

\begin{abstract}

We report the emergence of \textit{Moiré droplet} in two-dimensional ultracold bosons subjected to a spin-dependent optical lattice, effectively realizing a twisted-bilayer configuration. We show that the droplet formation dramatically enhances the visibility of Moiré pattern in the density profile, even for exceptionally weak lattice potentials. The Moire pattern can be  enhanced similarly by increasing the lattice depth, which, however, also induces  droplet diffusion characterized by a spreading density profile. Furthermore, we demonstrate a dynamical generation of Moiré pattern by dragging a small droplet through a moving lattice. At appropriate velocities, the droplet undergoes bifurcation and exhibits pronounced Moiré pattern within periodic time intervals. Our results establish the ultracold droplet as a compelling platform for simulating interacting Moiré physics, particularly the interplay between Moiré lattice and bound-state formation.

\end{abstract}
\maketitle

Moiré pattern describes a beautiful rippling structure generated by large-scale wave interference when two similar grids are overlaid with a relative rotation angle. In physics, it can give rise to new band structures that, when combined with interactions, lead to novel phenomena such as correlated insulators\cite{Insulator1, Insulator2} and unconventional superfluids\cite{SF1, SF2, SF3} as observed in twisted-bilayer graphene. Beyond two-dimensional (2D) materials, Moiré pattern has also been realized in photonic\cite{Photonic1,Photonic2, Photonic3, Photonic4} and ultracold atomic\cite{TBOL_expt} systems. Owing to its high controllability, the twisted-bilayer optical lattice in ultracold atoms has emerged as an ideal platform for engineering Moiré physics, in which atomic spin states serve as artificial layers and Rabi coupling between them provides the inter-layer tunneling\cite{proposal1, proposal2}. Theoretical studies of this system have revealed intriguing single-particle physics\cite{Santos1, Shi1, Shi2} and a variety of superfluid and insulating phases in repulsively interacting Bose gases\cite{He, Liubo, Lin, Santos2}.

When the interaction is switched from repulsive to attractive, bosons can organize themselves into an intriguing self-bound droplet. As recently realized in both dipolar gases\cite{Pfau_PRL2016, Pfau_Nature2016, Ferlaino_PRX2016, Modugno_PRL2019, Pfau_PRX2019, Ferlaino_PRX2019} and alkali mixtures\cite{Tarruell_Science2018,Tarruell_PRL2018,Fattori_PRL2018,Fort_PRR2019, Fort_CM2020, Wang_PRR2021}, the ultracold droplet is stabilized by a mean-field attraction arising from strong inter-species attraction and a beyond-mean-field repulsion originating from quantum fluctuations, which is therefore termed a quantum droplet\cite{Petrov}. Owing to its self-bound nature, the droplet exhibits unique collective excitations\cite{Petrov, Petrov_2, Liu, Reimann_1, Citro, Zhang_1, Ma_2} and dynamical behaviors\cite{Malomed_2, Fattori_2, Boronat, Modugno_3, Cui_3, Schmelcher} that are absent in conventional repulsive gases. In the presence of a twisted-bilayer lattice, an immediate question arises: how does the droplet respond to the underlying Moiré lattice? In particular, can a new phase emerge from the interplay between the self-bound structure and Moiré physics?

\begin{figure}[t]
    \centering
   \includegraphics[width=8.5cm]{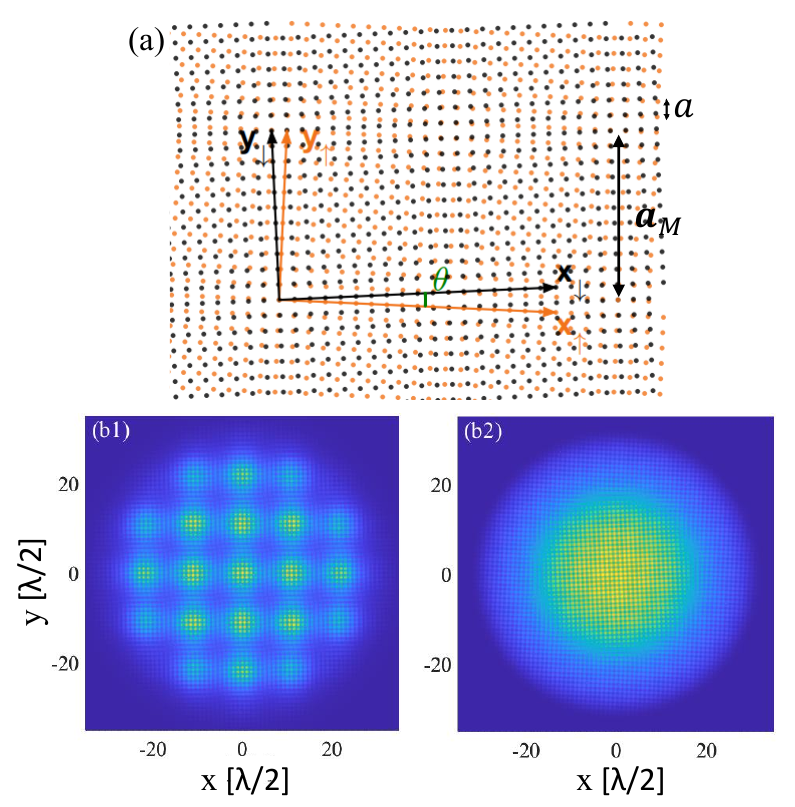}
    \caption{(a) Twisted-bilayer optical lattice. Orange and black dots respectively indicate the minima of lattice potential for $\uparrow$ and $\downarrow$ states. The twisted angle between the two lattices is taken as $\theta=5.21^{\circ}$, leading to the Moiré period $a_M= 11a$ ($a$ is the lattice spacing). (b1,b2) Density profiles of a quantum droplet (b1) and a repulsive gas (b2) under the potential in (a). The lattice depth is $V_0=E_R$. 
     For (b1), we take the realistic  $^{39}$K  system with $(a_{\uparrow\uparrow}, a_{\uparrow\downarrow}, a_{\downarrow\downarrow})=(73,-53,35)a_B$ (giving $\delta a=-2.45a_B$).  For (b2), we assume the same $a_{\uparrow\uparrow}$ and $a_{\downarrow\downarrow}$ as in (b1), but a different $a_{\uparrow\downarrow}=48a_B$; moreover, an additional harmonic trap with frequency $\omega_{\perp}=(2\pi)195$Hz is applied to avoid substantial spreading of the repulsive gas. The atom numbers in (b1,b2) are $N_{\uparrow}=28884$, $N_{\downarrow}=20000$.}\label{fig_lattice}
\end{figure}

In this work, we demonstrate a new self-bound state exhibiting Moiré pattern, which we term the \textit{Moiré droplet}. Specifically, we consider two-species bosons in 2D subject to a spin-dependent optical lattice, which effectively creates a twisted-bilayer system, see Fig.\ref{fig_lattice}(a). Here, the inter-species attraction replaces the Rabi field in \cite{TBOL_expt} to couple the two layers, and meanwhile, it facilitates the formation of a self-bound droplet in this system. In this way, the droplet density automatically captures the Moiré pattern of the underlying lattice. Remarkably, we find that droplet formation greatly enhances the visibility of Moiré pattern in the density profile, even under exceptionally weak lattice potentials. This is in sharp contrast to a repulsive gas in the same setup, as compared in Fig.\ref{fig_lattice}(b1,b2). 
Another route to enhancing the Moire pattern is to increase the lattice depth, which, however, also induces  droplet diffusion characterized by a spreading density profile.
Furthermore, we show that the Moiré pattern can be generated dynamically by dragging a small droplet through a moving lattice. At appropriate velocities, the droplet undergoes a periodic evolution between bifurcation and unification, with the former exhibiting clear Moiré pattern within periodic time intervals. The corresponding dynamical phase diagram is mapped out in the parameter plane of moving velocity and interaction strength. Our results reveal an intriguing interplay between quantum droplets and Moiré physics, which can be readily tested in current cold-atom systems.

We start from the energy functional of our system: ($\hbar=1$)
\begin{equation}
{\cal E}(\bm{\rho})= \sum_{\sigma} \psi_{\sigma}^* \left( -\frac{\nabla_{\bm{\rho}}^2}{2m} + V_{\sigma} \right) \psi_{\sigma} + \sum_{\sigma,\sigma'} \frac{g_{\sigma\sigma'}}{2} n_{\sigma}n_{\sigma'} + \varepsilon_{\rm LHY}. \label{H}
\end{equation}
Here $\bm{\rho}=(x,y)$ is the 2D coordinate and $m$ is the mass; $\psi_{\sigma}(\bm{\rho})$ is the spin-${\sigma}$ wavefunction and $n_{\sigma}=|\psi_{\sigma}|^2$ is its density; $g_{\sigma\sigma'}$ is the mean-field coupling between $\sigma$ and $\sigma'$ species, as given by $g_{\sigma\sigma'}= 4\pi a_{\sigma\sigma'}/(m\sqrt{2\pi}l_z)$ where $a_{\sigma\sigma'}$ is the 3D scattering length and $l_z=(m\omega_z)^{-1/2}$ is the characteristic length of longitudinal harmonic trap with frequency $\omega_z$. Near the mean-field critical point $g_{\uparrow\downarrow}^2\approx g_{\uparrow\uparrow}g_{\downarrow\downarrow}$ and under the local density approximation, the Lee-Huang-Yang (LHY) energy density can be written as\cite{supple}
\begin{equation}
\varepsilon_{\rm LHY} = \frac{m}{8\pi} f^2 \left[ \ln \left( \frac{fml_z^2\pi}{B} \right) + \frac{1}{2} \right], \label{E_LHY}
\end{equation}
with $f\equiv g_{\uparrow\uparrow}n_{\uparrow} + g_{\downarrow\downarrow}n_{\downarrow}$ and  $B=0.905$. In deriving (\ref{E_LHY}), we have combined the LHY correction in pure 2D\cite{Petrov_low_D} and the reduced scattering parameters in quasi-2D system\cite{Petrov_2D}. To validate Eq.(\ref{E_LHY}), one has to ensure a small value of $\eta\equiv f/\omega_z$\cite{Zin}, i.e., the characteristic mean-field energy is weak enough as compared to the longitudinal excitation gap. 
This imposes a constraint on the overall mean-field attraction, as parametrized by $\delta a\equiv a_{\uparrow\downarrow}+\sqrt{a_{\uparrow\uparrow}a_{\downarrow\downarrow}}$. To give an analytical estimation on $\eta$, we assume a homogeneous 2D droplet with zero pressure, from which we extract the equilibrium densities  and finally obtain $\eta$ as\cite{supple}
\begin{equation}
\eta=\frac{B}{\pi}\exp\left( \frac{\sqrt{8\pi}l_z }{(\sqrt{a_{\uparrow\uparrow}}+\sqrt{a_{\downarrow\downarrow}})^2} \frac{(-\delta a)}{\sqrt{a_{\uparrow\uparrow}a_{\downarrow\downarrow}}} -\frac{3}{2}\right). \label{eta}
\end{equation}
Clearly, the condition of small  $\eta$ precludes the system entering the strongly attractive regime, i.e., with large negative $\delta a$, where a dense droplet exists and gives rise to a large $\eta$.

Treating each spin state as a synthetic layer, we write down the twisted-bilayer lattice potential: 
\begin{eqnarray}
V_{\sigma}(\bm{\rho})&=&V_0  \sin^2 \left( kx\cos\frac{\theta}{2} - s_{\sigma}ky\sin\frac{\theta}{2} \right)  \nonumber\\
&+& V_0\sin^2 \left( ky\cos\frac{\theta}{2} + s_{\sigma}kx\sin\frac{\theta}{2} \right), \label{V_0}
\end{eqnarray}
where $s_{\updownarrow}=\pm 1$ and $\theta$ is the twisted angle between $V_{\uparrow}$ and $V_{\downarrow}$; the  wave-vector $k=2\pi/{\lambda}$ ($\lambda$ is the laser wavelength), giving lattice spacing $a=\lambda/2$ and recoil energy $E_R=k^2/(2m)$. In Fig.\ref{fig_lattice}(a), we take a small $\theta$ and plot out the  potential minima of $V_{\uparrow}$ and $V_{\downarrow}$, from which one can see the Moiré pattern with period $a_M=a/(2\sin(\theta/2))$. 

Based on (\ref{H}), we arrive at an extended Gross-Pitaevskii (GP) equation
\begin{equation}
i\frac{\partial\psi_{\sigma}}{\partial t} = \left( -\frac{\nabla_{\bm{\rho}}^2}{2m} + V_{\sigma} \right) \psi_{\sigma} + \sum_{\sigma'} g_{\sigma\sigma'} n_{\sigma'} + \frac{\partial\varepsilon_{\rm LHY}}{\partial n_{\sigma}} \psi_{\sigma}. \label{GP}
\end{equation}
The ground state can be obtained through an imaginary-time evolution of (\ref{GP}). In our simulation, we have taken the realistic system of $^{39}$K atoms with two hyperfine states $|\uparrow\rangle\equiv|F=1,m_F=-1\rangle,\ |\downarrow\rangle\equiv|F=1,m_F=0\rangle$, which can support  a quantum droplet near $B\sim57$G\cite{Tarruell_Science2018,Tarruell_PRL2018,Fattori_PRL2018}. Explicitly, we have  $a_{\uparrow\uparrow}=35a_B,\ a_{\uparrow\downarrow}=-53a_B$ ($a_B$ is the Bohr radius), and $a_{\downarrow\downarrow}$ is tunable by magnetic field. Throughout this work, we take the longitudinal trap length as $l_z=80nm$ and focus on the mean-field collapse regime with $73a_B<a_{\downarrow\downarrow}<a_{\downarrow\downarrow}^{(c)}=80a_B$, corresponding to $-2.45a_B<\delta a<0$ and $0.37>\eta>0.064$. Moreover,  we consider the tune-out wavelength $\lambda=769nm$ and thus the lattice spacing is $a=384.5nm$; the twisted angle is taken as $\theta=5.21^{\circ}$, giving the Moiré period $a_M=11a$.

Fig.\ref{fig_lattice}(b1) shows a typical ground-state density profile obtained from Eq.(\ref{GP}). The state is self-bound and exhibits a clear Moiré pattern that faithfully reflects the structure of the underlying Moiré lattice, and is therefore termed a \textit{Moiré droplet}. Remarkably, the lattice potential here is exceedingly shallow (with depth $V_0=E_R$), yet a visible  Moiré pattern still emerges in the droplet density profile. In contrast, for a purely repulsive Bose gas at the same  $V_0$, as shown in Fig.\ref{fig_lattice}(b2), no visible Moiré signature appears; instead, the whole cloud remains spatially symmetric, with no imprint of the Moiré lattice. 

The distinct behaviors in Fig.\ref{fig_lattice}(b1, b2) imply that  droplet formation can greatly enhance the visibility of Moiré pattern in the density of interacting bosons. Physically, this arises because the strong inter-species attraction in the droplet particularly favors a large density overlap between $\uparrow$ and $\downarrow$ atoms. Consequently, more atoms are driven toward the co-minima of the lattice potentials to maximize their density overlap, while fewer reside in the staggered regions, producing a clear Moiré pattern in the density distribution that fully respects the underlying lattice structure (see Fig.\ref{fig_lattice}(a)). In this way, the Moiré pattern is greatly enhanced by the attractive force, even in the presence of a very shallow lattice potential. This is in sharp contrast to the purely repulsive case, in which the shallow Moiré lattice is insufficient to support a clear Moiré pattern.


\begin{figure}[t]
    \centering
   \includegraphics[width=8.5cm]{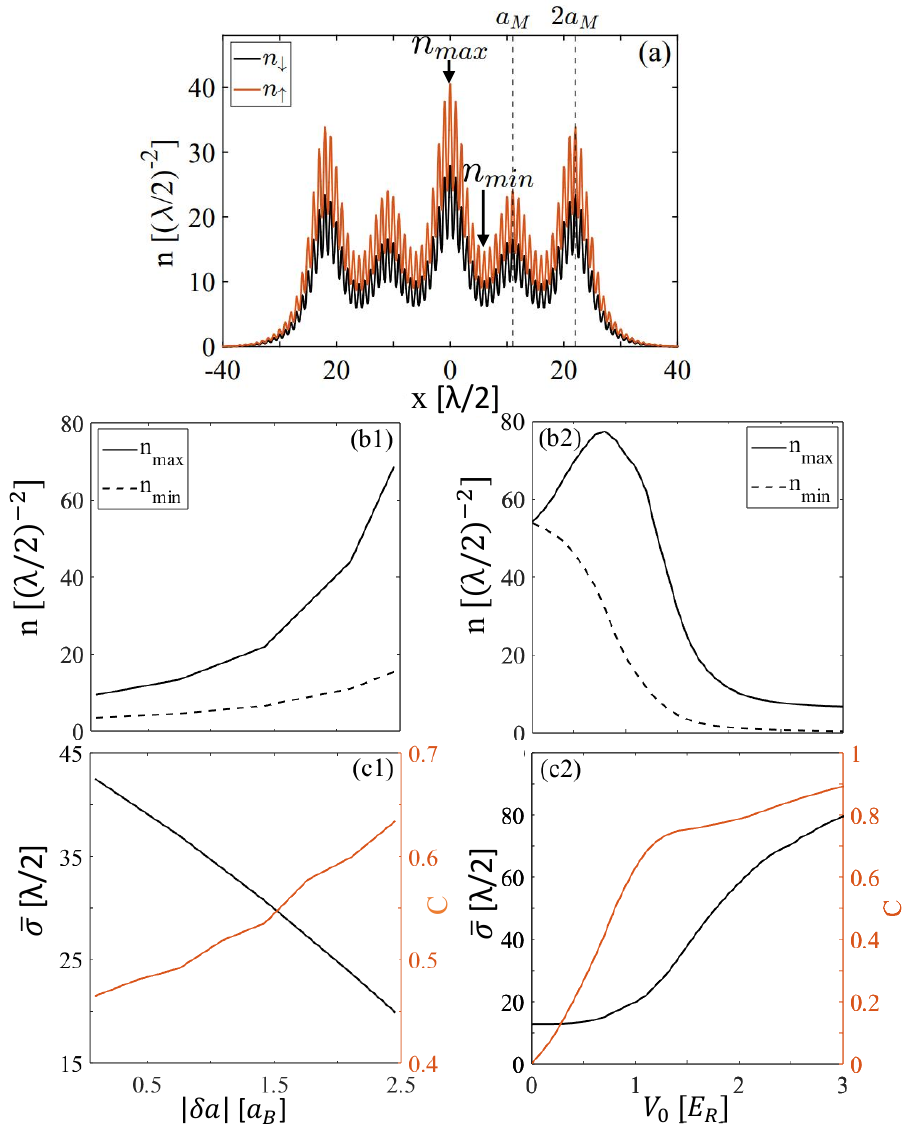}
    \caption{(a) Slice of density profile at $y = 0$, with $V_0=E_R$ and $\delta a=-2.45$. Here $n_{\rm max}$ and $n_{\rm min}$ respectively denote the total density maximum and minimum within a unit Moiré cell at the center.  (b1,c1) $n_{\rm max}, n_{\rm min}$, contrast $C$ and mean size $\bar{\sigma}$ as functions of $|\delta a|$ at a fixed $V_0=E_R$. (b2, c2) The same as (b1,c1) except as functions of $V_0$ at a fixed $|\delta a|=2.45a_B$. The atom numbers are the same as in Fig.\ref{fig_lattice}. }\label{fig_contrast}
\end{figure}

To parametrize the visibility of Moiré pattern, we define the contrast
\begin{equation}
C=\frac{n_{\rm max}-n_{\rm min}}{n_{\rm max}+n_{\rm min}},
\end{equation}
where $n_{\rm max}$ ($n_{\rm min}$) denotes the total density maximum (minimum) within a unit Moiré cell. As shown in Fig.\ref{fig_contrast}(a), we extract $n_{\rm max}$ and $n_{\rm min}$ from the Moiré cell at the center of the cloud. Their dependences on the attraction strength ($|\delta a|$) and the lattice depth ($V_0$) are shown in Fig.\ref{fig_contrast}(b1) and (b2), respectively, and the according  contrast ($C$) and mean size ($\bar{\sigma}\equiv \sqrt{\langle \bm{\rho}^2\rangle}$) are shown in Fig.\ref{fig_contrast}(c1,c2). We observe that $C$ grows continuously with increasing $|\delta a|$ or $V_0$. Nevertheless, increasing $|\delta a|$ and $V_0$ have markedly different effects on the fate of the droplet, as discussed below.

From Fig.\ref{fig_contrast}(b1,c1), we see that increasing $|\delta a|$ leads to tighter binding of the droplet, with higher $n_{\rm max}$ and $n_{\rm min}$ and a reduced $\bar{\sigma}$. In comparison, as shown in Fig.\ref{fig_contrast}(b2,c2),  increasing $V_0$ gives rise to a  non-monotonic change of $n_{\rm max}$, while $n_{\rm min}$ continuously decreases  and $\bar{\sigma}$ increases. These trends suggest an eventually diffusing droplet as increasing $V_0$, in contrast to the shrinking behavior as increasing $|\delta a|$. Such $V_0$-induced diffusion can be attributed to a larger mismatch between $V_{\uparrow}$ and $V_{\downarrow}$, which leads to severe spatial separation between $\uparrow$ and $\downarrow$ atoms and effectively reduces their attraction strength. As a result, the droplet becomes less bound and begins to diffuse. We emphasize that this behavior is unique to Moiré droplet and does not apply to the usual case without lattice twisting.    


In above we have demonstrated the Moiré droplet as the ground state of the system. In the following, we show that it can  equally be engineered dynamically by dragging a small droplet through a moving lattice. 
Here the moving lattice is created by introducing a frequency difference ($\Delta \omega$) between one pair of lasers acting on a spin state\cite{MovingLattice_review}. 
Specifically, we choose the $\uparrow$ state and modify the first term of $V_{\uparrow}$ as $V_0  \sin^2\left( kx\cos\frac{\theta}{2} - ky\sin\frac{\theta}{2} -kvt \right)$, with $v=\Delta \omega/(2k)$ the moving velocity. Owing to the interference between $V_{\uparrow}$ and $V_{\downarrow}$, this leads to a rapid motion of Moiré pattern along $y$ with velocity\cite{supple}
\begin{equation}
v_M=\frac{v}{2\sin(\theta/2)}. \label{v_M}
\end{equation}
Next we study the response of a quantum droplet to a moving Moire lattice.

\begin{figure}[t]
    \centering
   \includegraphics[width=8.5cm]{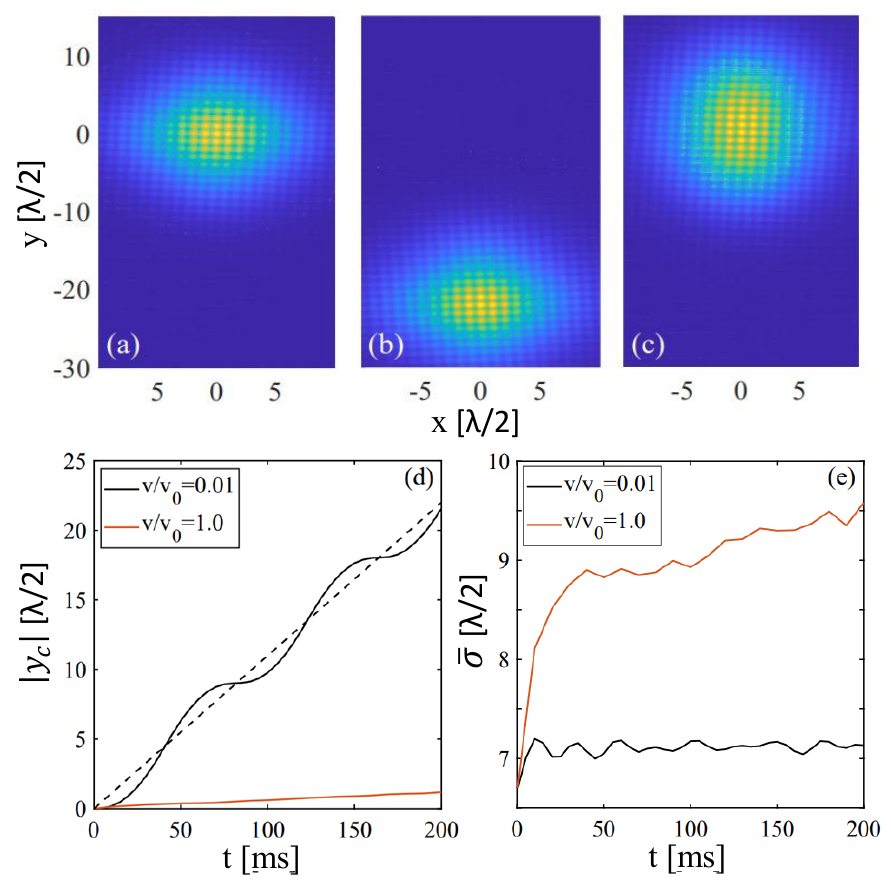}
    \caption{Time evolution of a small droplet under a moving Moiré lattice. (a-c)  density profiles of the initial state (a) and the droplet after time $t=200ms$ with different velocities $v/v_0=0.01 (b), 1(c)$ (here the unit $v_0=0.4mm/s$). (d) Time evolution of the center-of-mass position towards $-y$ direction (denoted by $|y_c|$) at different velocities. Dashed line shows linear fit to $|y_c|=v_Mt$, with $v_M$ given by (\ref{v_M}). (e) Time evolution of the droplet mean size $\bar{\sigma}$ at different velocities. Here $V_0=0.6E_R$, $\delta a=-2.45a_B$, and the atom numbers are $N_{\uparrow}=3610$, $N_{\downarrow}=2500$. }\label{fig_moving1}
\end{figure}

To begin with, we take the initial state as a small droplet within a unit Moire cell, as shown in Fig.\ref{fig_moving1}(a). Its dynamics is then governed by Eq.(\ref{GP}) with a moving $V_{\uparrow}$. Interestingly, we find the dynamical outcome   depends sensitively on the velocity $v$. For a very small $v$ (Fig.\ref{fig_moving1}(b)), the droplet adiabatically follow the underlying moving lattice,  with a linearly changing center-of-mass position $|y_c|=v_Mt$ (Fig.\ref{fig_moving1}(d)) and a saturated $\bar{\sigma}$ at long time (Fig.\ref{fig_moving1}(e)). In contrast, for a very large $v$ (Fig.\ref{fig_moving1}(c)), the droplet fails to follow the rapid motion of Moiré lattice, as manifested by the exceedingly slow change of $y_c$ in  Fig.\ref{fig_moving1}(d). During this process,  the droplet gradually expands with a growing $\bar{\sigma}$, see Fig.\ref{fig_moving1}(c, e); such diffusion can be understood as a consequence of  strong disorder caused by the rapidly moving lattice.

\begin{figure}[t]
    \centering
   \includegraphics[width=8.5cm]{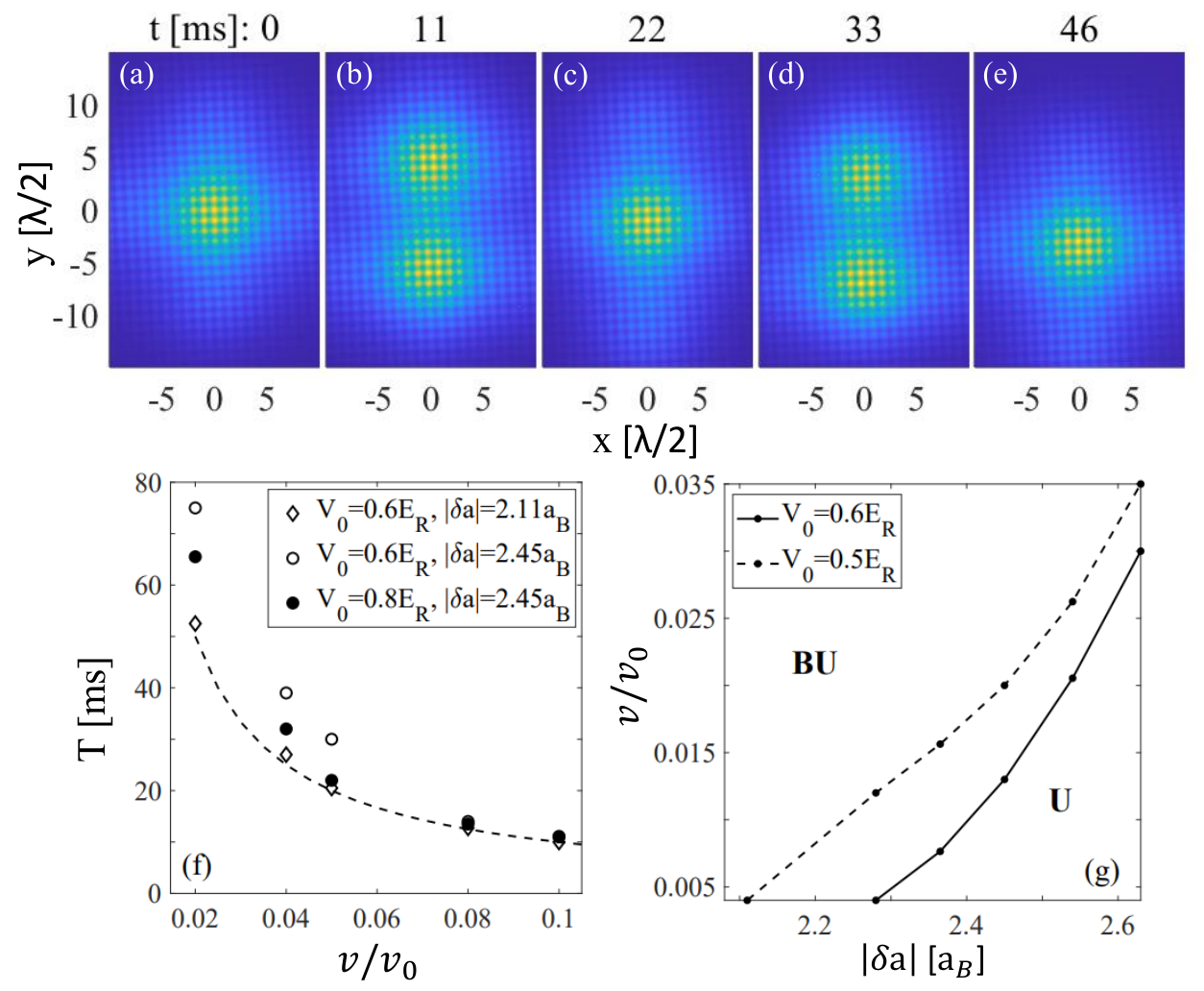}
    \caption{Droplet bifurcation dynamics under a moving Moiré lattice. (a-e) Time-dependent density profile  showing  periodic evolution between bifurcation (b, d) and unification (a, c, e). Here the moving velocity is $v/v_0=0.05$ with  $v_0=0.4mm/s$. (f) Period $T$ as a function of $v$ for different $\delta a$ and $V_0$. Dashed line shows fitting to $T=a/v$. (g) Dynamical phase diagram in the parameter plane of $|\delta a|$ and $v$ at different $V_0/E_R=0.5, 0.6$. The atom numbers are the same as in Fig.\ref{fig_moving1}.}\label{fig_moving2}
\end{figure}

The most intriguing situation arises at intermediate  $v$, where the droplet undergoes periodic evolution between bifurcation and unification, as shown in Fig.\ref{fig_moving2}(a-e). In the bifurcation phase (Fig.\ref{fig_moving2}(b,d)), the droplet splits into two along the moving ($y$) direction, with the two fragments separated by exactly one Moiré period  $a_M$. In this way, a Moiré droplet is created dynamically from the bifurcation of a single droplet. Such a bifurcation is driven by the center mismatch between the droplet and the time-dependent Moiré cell, and therefore can only persist  for a finite time interval. After a while, the system loses the Moiré feature and returns to a unified single droplet, as shown in Fig.\ref{fig_moving2}(c,e). To quantify this dynamics, we extract the period $T$ by monitoring the time duration over which the two split fragments in the bifurcation phase recover the same density peak. In Fig.\ref{fig_moving2}(f), $T$ is shown as a function of $v$ for various $|\delta a|$ and $V_0$. We find that $T$ generally decreases with $v$, and for weak attractions (small $|\delta a|$) and shallow lattices (small $V_0$) it approaches the ideal scaling $T=a_M/v_M=a/v$, see dashed line. For stronger attractions or deeper lattices, $T$ becomes visibly longer than the ideal prediction owing to stronger non-linear effects in this system. 


Physically, the periodic bifurcation-unification dynamics is a direct consequence of the intrinsic competition between underlying moving lattice ($V_0, v$) and droplet binding ($|\delta a|$). To identify their individual roles, in Fig.\ref{fig_moving2}(g) we map out a dynamical phase diagram in $|\delta a|$-$v$ parameter plane.  The diagram comprises two dynamical phases: one exhibiting periodic bifurcation-unification ('BU') evolution, and the other displaying only a single unified entity ('U'). In particular, we find that 'BU' is favored by a less bound droplet (smaller $|\delta a|$), a deeper Moiré lattice (larger $V_0$), and an intermediate $v$. Under these conditions, the moving lattice more readily tears the droplet into two and generates a Moiré droplet.

In summary,  we have demonstrated a new self-bound object, termed the Moiré droplet, for ultracold bosons in a twisted-bilayer optical lattice, which emerges either  as the ground state or as a dynamical outcome in a moving lattice. 
Our findings establish the ultracold droplet as a compelling platform for quantum simulation of interacting Moiré physics, particularly the interplay between the Moiré lattice and bound-state formation. This system opens avenues for exploring a wealth of intriguing quantum phenomena in future studies. For instance, as $V_0$ increases, the droplet may undergo a transition to a Moiré insulator, in which atoms become localized within individual Moiré cells due to the strong pinning effect of the twisted potential. Such a superfluid–insulator transition embodies both the Moiré characteristics and the self-bound nature of the droplet, and thus differs fundamentally from the conventional transition of repulsive bosons in optical lattices. Furthermore, relaxing the longitudinal trap would drive a dimensional crossover of the Moiré droplet from 2D to 3D, where a liquid-gas transition may appear due to the droplet diffusion in a deep Moiré lattice.
Finally, it would be interesting to investigate droplet behavior in an incommensurate Moire lattice with quasi-crystalline stacking order, which can host intriguing phenomena such as localizations and phason excitions\cite{NC2025,Science2025}.


\acknowledgments
This work is supported by National Natural Science Foundation of China (92476104, 12134015) and Innovation Program for Quantum Science and Technology (2024ZD0300600).

\clearpage

\onecolumngrid
\vspace*{1cm}
\begin{center}
{\large\bfseries Supplementary Materials}
\end{center}
\setcounter{figure}{0}
\setcounter{equation}{0}
\renewcommand{\figurename}{Fig.}
\renewcommand{\thefigure}{S\arabic{figure}}
\renewcommand{\theequation}{S\arabic{equation}}

In this Supplemental Material, we provide details on the derivation of Eqs.(2,3,7) in the main text.

\section*{I.\ \ \ Lee-Huang-Yang correction in quasi-2D}

As shown in Ref.\cite{Petrov_low_D}, near the mean-field collapse point the total energy density of 2D bosons can be written as
\begin{equation}
{\cal E}=\frac{1}{2}\sum_{\sigma\sigma'} g_{\sigma\sigma'} n_{\sigma}n_{\sigma'} +\frac{m}{8\pi} f^2 \left(\ln\frac{f}{\Delta}+\frac{1}{2}\right). \label{E_tot}
\end{equation}
Here $f=g_{\uparrow\uparrow}n_{\uparrow}+g_{\downarrow\downarrow}n_{\downarrow}$; $\Delta$ is an energy parameter that can be related to the effective 2D  coupling via 
\begin{equation}
g_{\sigma\sigma'}=-\frac{4\pi}{m\ln(ma^2_{2D;\sigma\sigma'}\Delta)}, \label{g}
\end{equation}
with  $a_{2D;\sigma\sigma'}$ the  2D scattering length in $\sigma-\sigma'$ scattering channel. 

For a realistic quasi-2D system with a longitudinal confinement,  $\Delta$ is of the same order as the characteristic energy  scale along the confined ($z$) direction. For instance, for a uniform confinement of length $l_z$, we have the 2D coupling $g_{\sigma\sigma'}=\frac{4\pi a_{\sigma\sigma'}}{ml_z}$ and $a_{2D;\sigma\sigma'}=l_ze^{-l_z/(2a_{\sigma\sigma'})}$\cite{Cui}, with $a_{\sigma\sigma'}$ the 3D scattering length in $\sigma-\sigma'$ scattering channel. Then based on (\ref{g}), we can obtain $\Delta=1/(ml_z^2)$, and the LHY correction, as given by the second term in (\ref{E_tot}), exactly reproduces the result in Ref.\cite{Zin} obtained by a different method.

For a harmonic confinement along $z$ with frequency $\omega_z$ and trap length $l_z=1/\sqrt{m\omega_z}$, we have 
$g_{\sigma\sigma'}=\frac{4\pi a_{\sigma\sigma'}}{\sqrt{2\pi}ml_z}$ and $a_{2D;\sigma\sigma'}=\sqrt{\pi}{B}l_ze^{-\sqrt{\pi/2}l_z/a_{\sigma\sigma'}}$ ($B=0.905$)\cite{Petrov_low_D}, and then from (\ref{g}) we obtain $\Delta=(B/\pi)\omega_z$ and the LHY energy density can be expressed by Eq.(2) in the main text.

\section*{II.\ \ \ Derivation of $\eta$ for a homogeneous 2D droplet}

Near the mean-field collapse point, we can express $g_{\uparrow\downarrow}=-\sqrt{g_{\uparrow\uparrow}g_{\downarrow\downarrow}}+\delta g$, with $|\delta g|\ll g_{\sigma\sigma}$. In this regime,  the total energy density in (\ref{E_tot}) can be simplified as
\begin{equation}
{\cal E}=\frac{1}{2} (\sqrt{g_{\uparrow\uparrow} }n_{\uparrow}-\sqrt{g_{\downarrow\downarrow} }n_{\downarrow})^2 + \delta g n_{\uparrow} n_{\downarrow} + \frac{m}{8\pi} f^2 \left(\ln\frac{\pi f}{B\omega_z}+\frac{1}{2}\right).
\end{equation}
 The minimization of mean-field part leads to a locked density ratio $n_{\uparrow}/n_{\downarrow}=\sqrt{g_{\downarrow\downarrow}/g_{\uparrow\uparrow}}$\cite{Petrov}. Using this relation, ${\cal E}$ can be expressed as a function of a single density $n_{\downarrow}$. Then using the zero-pressure requirement for a homogeneous droplet, i.e., $-P={\cal E}-n_{\downarrow}\partial {\cal E}/\partial n_{\downarrow}=0$, we can uniquely solve $n_{\downarrow}$ as
 \begin{equation}
 n_{\downarrow}=\frac{B\omega_z}{\pi g_{\downarrow\downarrow}}\frac{1}{1+\sqrt{a_{\uparrow\uparrow}/a_{\downarrow\downarrow}}} \exp\left( \frac{\sqrt{8\pi}l_z }{(\sqrt{a_{\uparrow\uparrow}}+\sqrt{a_{\downarrow\downarrow}})^2} \frac{(-\delta a)}{\sqrt{a_{\uparrow\uparrow}a_{\downarrow\downarrow}}} -\frac{3}{2}\right).
 \end{equation}
 Further, $n_{\uparrow}$ can be obtained according to the locked ratio. Finally, we can get the parameter $\eta\equiv f/\omega_z$ as Eq.(3) in the main text. 
 
\section*{III.\ \ \ Moving velocity of a Moiré cell }

The moving lattice we consider in this work is only acted on spin-$\uparrow$ atom, with potential
\begin{eqnarray}
V_{\uparrow}&=&V_0  \sin^2 \left( kx\cos\frac{\theta}{2} - ky\sin\frac{\theta}{2} -kvt \right) + V_0\sin^2 \left( ky\cos\frac{\theta}{2} + kx\sin\frac{\theta}{2} \right), \label{V_up}
\end{eqnarray} 
while for spin-$\downarrow$, the lattice potential is kept unchanged:
\begin{eqnarray}
V_{\downarrow}&=&V_0  \sin^2 \left( kx\cos\frac{\theta}{2} + ky\sin\frac{\theta}{2}  \right) + V_0\sin^2 \left( ky\cos\frac{\theta}{2} - kx\sin\frac{\theta}{2} \right). \label{V_down}
\end{eqnarray} 
The movement of Moiré cell can be examined from the total potential $V_{\uparrow}+V_{\downarrow}$, which reads
\begin{eqnarray}
V_{\uparrow}+V_{\downarrow}&=&-V_0  \cos \left( 2kx\cos\frac{\theta}{2} - kvt \right) \cos \left( 2ky\sin\frac{\theta}{2} + kvt \right) -V_0 \cos \left( 2kx\sin\frac{\theta}{2}\right) \cos \left( 2ky\cos\frac{\theta}{2} \right) . \label{V_t}
\end{eqnarray}
Let us follow the trajectory of a Moiré cell centered at a local minimum of $V_{\uparrow}+V_{\downarrow}$. At initial time $t=0$, we consider a local minimum at $(x,y)=(0,0)$; at time $t$, such a minimum moves to $(vt/(2\cos\frac{\theta}{2}), vt/(2\sin\frac{\theta}{2}))$ according to the first term of (\ref{V_t}). For a small $\theta$, we can see that the motion of the Moiré cell along $x$ is much slower than that along $y$, and meanwhile, such a minimum also approximately represents  the minimum of the second term in (\ref{V_t}). We can therefore extract the moving velocity of the minimum as $v_M=v/(2\sin\frac{\theta}{2})$  along $y$ (i.e., Eq.7 in the main text), while the motion along $x$ can be negligible. Indeed we have  confirmed this conclusion in our numerical simulation of the moving Moiré lattice. 

Remarkably, although the moving lattice is generated by a pair of lasers oriented essentially along $x$ direction (see Eq.\ref{V_up}), the resulting motion  of the Moiré pattern is essentially along $y$, which is orthogonal to the moving direction of original $\uparrow$-lattice. This counterintuitive result can be attributed to the non-trivial interference between $\uparrow$ and $\downarrow$ lattices in creating the Moiré pattern. It also suggests that the motion of the Moiré pattern can be conveniently controlled by moving the lattices for different spins (layers) or along different directions.

\end{document}